**Raman fingerprints of spin-phonon coupling and magnetic transition in an organic molecule intercalated Cr₂Ge₂Te₆**

Sudeshna Samanta[1], Hector Iturriaga[1], Thuc T. Mai[2], Adam J. Biacchi[3], Rajbul Islam[4], Angela R. Hight Walker[2], Mohamed Fathi Sanad[5], Charudatta Phatak[6], Ryan Siebenaller[7,8], Emmanuel Rowe[7,9,10,11,12] Michael A. Susner[7], Fei Xue[4], Srinivasa R. Singamaneni[1]*

[1]Department of Physics, The University of Texas at El Paso, El Paso, Texas 79968, USA

[2]Quantum Metrology Division, Physical Measurement Laboratory, National Institute of Standards and Technology, Gaithersburg, Maryland 20899, USA

[3]Nanoscale Device Characterization Division, Physical Measurement Laboratory, National Institute of Standards and Technology, Gaithersburg, Maryland 20899, USA

[4]Department of Physics, University of Alabama at Birmingham, Birmingham, AL 35233

[5]Department of Chemical Engineering, Hampton University, Hampton, VA 23668, USA

[6]Materials Science Division, Argonne National Laboratory, Lemont, IL, USA.

[7]Materials and Manufacturing Directorate, Air Force Research Laboratory, Wright-Patterson Air Force Base, OH 45433 USA.

[8]Department of Materials Science and Engineering, The Ohio State University, Columbus, OH 43210 USA

[9]National Research Council, Washington, DC 20001, USA

[10]Department of Engineering Technology, Middle Tennessee State University, Murfreesboro, TN 37132, USA

[11]Department of Astronomy and Physics, Vanderbilt University, Nashville, TN 37235, USA

[12]Department of Life and Physical Sciences, Fisk University, Nashville, TN 37208, USA

**Abstract**

The manipulation of spin-phonon coupling in both formations and explorations of magnetism in two-dimensional van der Waals ferromagnetic semiconductors facilitates unprecedented prospects for spintronics devices. The interlayer engineering tunes spin-phonon coupling significantly and holds the promise for controllable magnetism via organic cation intercalation. Here, we present spectroscopic evidence to reveal the intercalation effect on intrinsic magnetic and electronic transitions in quasi-two-dimensional Cr₂Ge₂Te₆ using tetrabutyl ammonium as the intercalant. The temperature-evolution of Raman modes $E_g^3$ and $A_g^1$, along with the magnetization measurements, unambiguously captures the enhancement of the ferromagnetic Curie temperature in the intercalated heterostructure. Moreover, the $E_g^4$ mode highlighted the increased effect of spin-phonon interaction in magnetic order-induced lattice distortion. Combined with the first-principle calculations, we observed a substantial number of electrons transferred from TBA$^+$ to Cr through the interface. These results provide the interplay between spin-phonon coupling and magnetic ordering in van der Waals magnets where Raman fingerprints would be highly beneficial for further understanding the manipulation of magnetism in layered heterostructures.

*srao@utep.edu

**Introduction**

Recent advances in the newly developing spintronics and valleytronics technology have deepened the search for magnetism in two-dimensional (2D) materials where charge and spin degrees of freedom of charge carriers control their intriguing properties. The recent discoveries of few-layer and monolayer 2D van der Waals (vdW) crystals such as $Cr_2Ge_2Te_6$ (CGT) [1, 2], $CrI_3$, and $Fe_3GeTe_2$ reveal the existence of long-range ferromagnetic (FM) order at low temperatures. Successful spintronics device fabrication depends on the effective control of their magnetism by tuning the interlayer magnetic coupling with carrier doping, strain, pressure [3, 4], electric field [5], and more recently, through cation intercalation [6] while achieving the room temperature ferromagnetism. The intercalation of foreign ions in the vdW magnet unwraps the possibility of manipulating magnetism and carrier dynamics by electron-phonon and spin-orbit coupling (SOC) mechanisms, respectively. Such coupling mechanisms are critical for devices' magnetic and thermal relaxation processes and explain other novel effects like multi-ferroelectricity. Additionally, the heat dissipation mechanism in prototype devices is crucial and involves a comprehensive understanding of the interplay between lattice vibrations (phonons) and magnetism. Such studies also help to explain the magneto-elastic effects for potential devices fabricated with 2D vdW magnets.

In this context, Raman spectroscopy is a truly appreciated technique containing information about intra-layer and inter-layer (phonon) modes stemming from chemical bonds (intra-layer) and layer-layer vdW interactions, respectively. The response of phonon modes to temperature exhibits several prominent features containing valuable information [7] about spin dynamics, electron-phonon coupling (EPC), phonon-anharmonicity, etc. Here, we probe the interfacial coupling mechanism and its close correlation with magnetic and electronic phase transitions in a quasi-2D vdW magnet and its intercalated heterostructure. The high-resolution Raman shifts can precisely record sub-cm$^{-1}$ changes in the vibrational frequencies entangled with their lattice dynamics.

A pristine $Cr_2Ge_2Te_6$ (CGT) is a ferromagnetic (FM) semiconductor with a band gap of $\approx$ 0.2 – 0.74 eV below the Curie temperature $T_C \approx$ 61-67 K [1] (**Fig. S1**). The $Cr^{3+}$ atoms carry itinerant spins ($S = 3/2$) and local FM moments and are octahedrally coordinated by Te ligands to form a 2D honeycomb lattice in $ab -$ plane. The individual magnetic layers are stacked and weakly coupled by vdW interactions along the $c -$ axis. Prior work carried out by other researchers on an organic cation (tetrabutyl ammonium: TBA$^+$) intercalated heterostructure (TBA-CGT) of CGT reported a dramatic elevation of $T_C \approx$ 208 K [6]. The magnetic easy axis flipped from $\langle 001 \rangle$ $c -$ axis in CGT to $ab -$ plane in TBA-CGT due to the transformation from weak magnetic superexchange interaction to a strong double-exchange interaction respectively [6]. Moreover, TBA-CGT turned into a metallic state at low temperatures, showing a semiconductor-metal transition around 165 K [6]. However, it remains elusive how the intercalation-induced magnetism is connected with spin-phonon coupling, which might have more significant implications for potential applications.

To that end, it is worth investigating the temperature ($T$) response of inter-layer Raman modes in TBA-CGT, especially as intercalation greatly influences the vdW interaction due to the modification in stacking order and change in magnetic response via exchange interactions. What differentiates our study from prior literature work [6] is the use of temperature-dependent Raman fingerprints to track the magnetic (ferromagnetic → paramagnetic) and electronic (semiconductor → metal) transitions in pristine and intercalated structures. Finally, the Raman spectroscopy and magnetization measurements provide a comprehensive platform to understand phonons with high spatial and spectral resolution. In parallel with the experimental realizations, theoretical simulations based on the first-principles calculations have also contributed significantly to understanding intercalated CGT's electronic and magnetic properties.

**Result and discussions**

The details related to the sample synthesis and experimental techniques are discussed under the experimental methods section. We studied the structural details and bonding characteristics of pristine and intercalated CGT crystals using X-ray diffraction (XRD) and X-ray photoelectron spectroscopy (XPS). These measurements confirmed the successful intercalation of TBA. We discussed about these findings in the supplementary information (**Fig. S2–S4**).

At zero temperature, the pristine CGT is described as a Heisenberg ferromagnet [8]. It was reported that the ferromagnetic transition temperature $T_C$ in 2D and 3D cases primarily depends on the excitation gap that originates from magnetic anisotropy and magnetic exchange interactions, respectively. Therefore, ion-mediated electrochemical intercalation is expected to alter CGT's inter-plane magnetic exchange interactions and provide an ideal platform to study the magnetic ground state and spin-phonon coupling (SPC) mechanism.

To observe the status of ferromagnetism in TBA-CGT, we plot the temperature-dependent magnetization ($M - T$) in **Fig. 1(a)** and **Fig. 1(b)** when the applied magnetic field is along $H//c$ axis and $H//ab$ plane, respectively. The derivative of magnetization ($dM/dT$) in **Fig. 1(a-I)** shows FM ordering Curie temperature $T_C^{in} \approx 204$ K (after TBA intercalation) and agrees well with 208 K reported earlier [6]. Strikingly, upon cooling, two consecutive minima in $dM/dT$ curve around $T_k \approx 178$ K and at the Curie temperature $T_C \approx 66$ K [6, 9] are distinguishable and are absent (see, **Fig. 1(b-I)**) when measured under $H//ab$ plane configuration. We note that $T_k \approx 178$ K was unreported in the literature for TBA-intercalated CGT [6], while $T_C \approx 66$ K represents the ferromagnetic ordering temperature of the pristine CGT.

The transition temperatures $T_C^{in}$ and $T_k$ in the intercalated system reflect the actual effect of ion migration and its impact on (quasi) 2D ferromagnetic order in our system. First, these temperatures depend purely on the crystal lattice because the ions do not bond strongly with CGT layers. Second, TBA$^+$ did not affect the intrinsic $T_C$, and the observation of $T_C^{in}$ is unambiguously strong evidence of a long-range secondary 2D ferromagnetism originating from the alteration of CGT atomic layers.

A strong effect of magnetic anisotropy is revealed by the isothermal magnetization ($M - H$ curves) measurements conducted at various temperatures as plotted in **Fig. 1(c)-(d)**. TBA-CGT possesses a typical FM behavior with a very low saturation magnetic field ($\approx 0.5$ Tesla) at the lowest temperature of 2 K when $H//c$ (easy axis). In contrast, when $H//ab$ (hard-axis), no saturation of the magnetization exists even at the highest magnetic field of 7 Tesla. With increasing temperature, the change in the curvature of $M - H$ loop indicates the onset of ferromagnetic-paramagnetic transition above 200 K. We calculated the magnetic anisotropy constant ($K_u$) for TBA-CGT and pristine CGT (**Fig. S5**) along the magnetic hard axis ($H//ab$) and it decreases significantly after intercalation. Moreover, we notice that the magnetic easy axis does not alter from $c-$ axis to $ab-$plane upon intercalation, in contrast with the earlier study [6]. It was shown earlier [5] that heavy doping (by electrostatic gating) switches the sign of the magnetic anisotropy energy and alters the magnetic easy axis from out of plane to in plane. However, in the present case, the easy axis of magnetization is stable and still preserves the long-range FM order with a high $T_c^{in} \approx 204$ K upon TBA intercalation. Similar to the present results, we recently reported that the magnetic easy axis did not alter in the TBA-intercalated Fe$_{3-x}$GeTe$_2$ [9].

Raman spectroscopy is a sensitive tool for analyzing long-range magnetic order and investigating the spin-phonon coupling in magnetic transitions in 2D magnetic materials. CGT possesses a rhombohedral crystal structure [2] as $R\overline{3}(C_{3i}^2)$ and the group theory analysis shows that there exist 10 Raman active modes as $\Gamma = 5E_g + 5A_g$ in pristine CGT [3, 10, 11] (Table S-I). **Fig. 2(a)** shows the Raman spectrum for TBA-CGT at 295 K (Table S-I), and the temperature evolution of Raman spectra measured from 1.6 K to 295 K is depicted in **Fig. 2(b)**. We track the behaviors of most intense modes $E_g^3$ and $A_g^1$ with temperature to explore the magnetic transitions in TBA-CGT. In pristine CGT, low-frequency ($< 150$ cm$^{-1}$), mid-frequency ($180 - 240$ cm$^{-1}$), and high-frequency ($> 260$ cm$^{-1}$) modes are dominated by motions of Te, Cr, and Ge atoms, respectively [3, 11]. The emergence of a low-frequency Raman inactive mode $A_{2u}$ around 58.9 cm$^{-1}$ might be a part of the Davydov doublet arising from the resonant effect [12].

A closer inspection of the two-dimensional intensity contour plots reveals non-linear temperature dependencies of $E_g^3$ and $A_g^1$ modes in **Fig. 2(c)-I** (TBA-CGT) and **Fig. 2(c)-II** (pristine CGT) respectively. These modes red-shift with the progression of temperature, and upon cooling, $E_g^3$ and $A_g^1$ display several corrugated or kink-like features at temperature scales $T_C^{in}$ ($\approx 204$ K), $T_k$ ($\approx 178$ K), $T_1$ ($\approx 102$ K) (discussed below), and $T_C$ ($\approx 66$ K) in TBA-CGT as depicted in **Fig. 2(c-I)**. Surprisingly, the above temperatures highly coincide with the temperature minima in $dM/dT$ curve in **Fig. 1(a-I)**. In contrast, in the pristine CGT, only one kink-like feature is present for both modes around $T_C$ ($\approx 66$ K) (**Fig. 2(c)-II**, **Fig. 1(b-I)**) where frequencies shift smoothly to a higher energy side. We also observed the rapid reduction in the intensity ratio of $E_g^3$ and $A_g^1$ modes with increasing temperature, suggesting the in-plane vibrations dominate over the out-of-plane vibrations (**Fig. 2(b)-(c)**). Therefore, fluctuations in the phonon mode intensities, along with the variations in slopes in vibrational frequencies, are associated with the magnetic orders and electronic properties of these 2D materials.

All the Raman modes in TBA-CGT moved to lower wave numbers (**Fig. 3(a)**) and confirmed the electron doping mechanism in CGT. We suggest that with TBA$^+$ intercalation, some of the Cr$^{3+}$ ions are replaced by Cr$^{2+}$ ions (local cation doping), and a double-exchange mechanism might be initiated. Upon cooling, hopping appears to be a more energetically favorable mechanism in semiconductors like CGT. As a result, the hopping across Cr$^{3+}$-Te-Cr$^{2+}$ links stabilizes the magnetic order. It preserves the electron spins in the system, and a remarkable semiconductor-metal transition [6] occurs around $T_k$ in sharp contrast to a semiconducting character in CGT. The first-order electronic transition in the TBA-CGT occurs by the carrier-mediated indirect exchange mechanism (via Cr$^{3+}$-Te-Cr$^{2+}$ links), where the charge transfer happens by localized electrons. Additionally, the considerable FM exchange interaction ($J_1$) arises from the competition between the negative direct exchange (Cr$^{3+}$- Cr$^{3+}$ sites) and the positive superexchange (Cr$^{3+}$-Te-Cr$^{3+}$) interactions [11]. Further, TBA$^+$ intercalation causes the lattice expansion along the $c-$axis and significantly weakens the interlayer nearest Cr's exchange interaction. As a result, a strong SPC in the CGT layers nearest to the TBA interfaces tunes the electrical transport and magnetic order simultaneously. $E_g^3$ and $A_g^1$ modes involve the intralayer Te motions around Cr atoms in $ab-$ plane during different magnetic exchange interactions and explain why they are highly susceptible to temperature. The red shift of the $E_g^3$ mode indicates the compression of the associated bonds. It coincides with a high-pressure study [13], suggesting that chemical pressure-induced intercalation modulates this vibration similarly. Here, we conclude that electron doping impacts the in-plane phonon vibrations, revealing a quasi-2D electron-phonon interaction above $T_1 \approx 102$ K [9, 14].

Raman frequencies also follow non-linear temperature dependencies (**Fig. 3(a)**), and it is shown for a representative $E_g^3$ mode in **Fig. 3(b)** when plotted simultaneously with its electrical resistance [6]. The slope changes at different rates (– 0.007 cm$^{-1}$/K, – 0.015 cm$^{-1}$/K, – 0.004 cm$^{-1}$/K, and – 0.017 cm$^{-1}$/K) at different temperature regimes encompassing the electronic and magnetic transitions. In contrast, $E_g^3$ mode for CGT remains less sensitive (-0.003 cm$^{-1}$/K for 100 – 300 K [3, 10]) or changes monotonically with temperature. The modes like $A_g^3$, $E_g^2$, and $E_g^1$ involving the motions of Te atoms around Cr atoms do not change substantially and are very robust against temperature. These modes blue-shift below $T_C^{in}$, therefore, these bonds are squeezed due to the intercalated cations.

As hinted above, the electrochemical cation intercalation-induced electron doping appears analogous to the electrostatic gating effect [5]. If the intercalation is fully realized, the high-density and localized electrons are available between monolayers. Such injection of additional electrons at the interfaces between two materials generally gives rise to a charge transfer between the layers. To assess this effect, we define the linear charge density $\rho_{eff}(z)$ as the number of electrons summed over the $xy-$ plane per unit length such that $\int \rho_{eff}(z)dz$ is equal to the number of electrons. We have performed density functional theory calculations for the entire heterostructure and separate slabs containing only CGT and TBA with the same geometry as the heterostructure. Subsequently, we have considered the difference between the respective linear charge densities:

$$\Delta\rho(z) = \rho_{CGT/TBA+}(z) - \rho_{CGT}(z) - \rho_{TBA+}(z) \tag{1}$$

The integral of this quantity is not zero by construction. We find that a substantial number of electrons from TBA$^+$ transfer to Cr through the interface, which gives rise to charge accumulation near the interfaces, as shown in **Fig. 4**. These results infer that there is significant charge transfer between TBA$^+$ and CGT in the superposition, which are consistent with our experimental findings. The calculation in the Bader charge analysis is depicted in Table I. Furthermore, we have calculated the electronic band structure of the CGT/TBA$^+$ heterostructure using experimental structures with an out-of-plane magnetization (ground state in the presence of SPC, it shows the presence of metallic bands at the Fermi level [see **Fig. 5** (left panel)], indicating a semiconductor-metal transition with intercalation. We see the metallic band contributions are mainly from the Cr$-d$ and Te$-p$ orbitals, as shown in the right panel of **Fig. 5**. Note that TBA$^+$ electronic bands are far from the Fermi level, as shown in **Fig. 5**.

**Table I.** The calculated net charge transfer of corresponding atoms in TBA-CGT superlattice.

| Atoms | Bottom-layer | Middle-layer | Top-layer |
|-------|--------------|--------------|-----------|
| Cr | +0.927 | +0.927 | +0.926 |
| Ge | +0.514 | +0.488 | +0.773 |
| Te | -0.650 | -0.830 | -0.573 |
| N | -1.993 | -2.006 | -1.985 |
| C | +0.211 | +0.503 | +0.143 |
| H | -0.005 | +0.019 | -0.075 |

The mechanism of magnetic transition depends on the strength of SPC via the involvement of the Cr motion in each vibrational frequency [11, 15]. Hence, $E_g^4$ and $E_g^2$ modes have the largest SPC, about 3.19 cm$^{-1}$ and 1.24 cm$^{-1}$, respectively [11]. The rest of the Raman modes have weak SPC coefficients due to the involvement of Ge or Te atoms. Interestingly, as $E_g^4$ mode involves the antiphase motions of the Cr atoms, the temperature dependency of the $E_g^4$ mode for both samples is qualitatively different (linear temperature dependency) from their own $E_g$ counterparts. $E_g^4$ mode in TBA-CGT softens faster (-0.034 cm$^{-1}$/K) than CGT (-0.011 cm$^{-1}$/K) (**Fig. 2(a)**). Theoretical calculations also show that the SPC coefficient for $E_g^4$ mode alters rapidly with lattice strain and is inversely proportional to the lattice constant [3]. Therefore, the onset of FM order and an increase of the SPC strength might soften $E_g^4$ mode rapidly below 204 K. It also explains the direct impact of intercalation on the temperature-dependent lattice shrinkage and magnetic order-induced lattice distortion.

Temperature evolution of Raman shifts can be expressed as $\omega\,(T) = \omega_0 + \Delta\omega_{vol}(T) + \Delta\omega_{anh}(T)$, where $\Delta\omega_{vol}(T)$ and $\Delta\omega_{anh}(T)$ are the volume/implicit and the true/explicit anharmonicity counterparts respectively [16]. We applied the extended Klemens-Hart-Aggarwal-Lax phonon decay model considering the three and four-phonon process into two (three phonon) and three (four phonon) acoustic phonons as

$$\omega(T) - \Delta\omega_{vol}(T) = \omega_0 + M_1\left[1 + \frac{2}{e^x-1}\right] + M_2\left[1 + \frac{3}{e^y-1} + \frac{3}{(e^y-1)^2}\right] \quad (2),$$

where $\omega_0$ is the frequency, $x = \hbar\omega_0/2k_BT$, $y = \hbar\omega_0/3k_BT$, $\hbar\omega_0$ is the energy of the optical phonon frequency $T = 0$, $\hbar$ is the Planck's constant, $k_B$ is the Boltzmann constant, and $M_1$ and $M_2$ are the anharmonic constants. The solid lines in **Fig. 6** show best fit (Table S-II) to the experimental data using Equation (2) for $T_C < T < T_C^{in}$ in TBA-CGT. The fitting demonstrates that the three-phonon process is dominating over the four-phonon process.

The modes in the paramagnetic phase $(T > T_C^{in})$ strongly deviate from the Boltzmann-sigmoidal fits. Thus, we estimate the frequency differences $\Delta\omega$ at the lowest temperature (0 K) and $T_C^{in}$ (see, **Fig. 6**). Our analysis shows that spin-phonon interaction parameters $\lambda'$ for $E_g^3$, $A_g^1$, and $E_g^4$ are three times higher than $\lambda'$ values for CGT [10]. $\lambda'$ is higher for $E_g^4$ compared to $E_g^3$ and $A_g^1$ in TBA-CGT and reconfirms the substantial modification of the exchange coupling after intercalation.

Here, we reveal a close correlation between EPC and a sharp drop of resistance [6] around $T_1 \approx 100$ K in TBA-CGT. The extracted $E_g^4$ mode linewidth as a function of temperature (see, **Fig. 7**) shows anomalous and distinctly nonlinear characteristics. The intrinsic linewidth depends on the phonon-phonon (lattice anharmonicity) and electron-phonon interaction. It can be expressed as $\Gamma(T) = \Gamma_0 + \Gamma_{anh}(T) + \Gamma_{EPC}(T)$ where $\Gamma_0$, $\Gamma_{anh}(T)$, $\Gamma_{EPC}(T)$ are temperature independent linewidth, linewidth due to anharmonicity in lattice vibrational potential, and contribution from EPC associated with Raman modes. $\Gamma_{anh}(T)$ can be expressed in three- and four-phonon processes as described above. In the presence of strong EPC, a rapid reduction in linewidth occurs below 100 K (see, **Fig. 7**). The EPC strength at absolute zero temperature is estimated as $\Gamma_{EPC}(0) \approx 3.72 \pm 0.31$ cm$^{-1}$. Around 100 K, the phonon lifetime is the longest, about 1.5 ps. Recently a THz spectroscopy study on CGT reported the highest center frequency, around 0.87-0.9 THz, corresponding to a phonon lifetime of around 1.15-1.11 ps [17]. They also observed an attenuation of FWHM due to the formation of magnetic correlations below the spin fluctuation temperature of 160 K [17]. However, $\tau$ decreases due to the increased phonon-phonon scattering events at higher temperatures. For the other modes, only phonon-phonon interaction exists without any EPC. We suggest that EPC might have triggered the semiconductor-metal transition in TBA-CGT, incorporating a sharp resistance drop around $T_1$.

The lifting of degeneracy due to time-reversal symmetry breaking leads to splitting the lowest energy phonon mode ($E_g^1$: 78.6 cm$^{-1}$) in CGT at low temperatures [10]. It originates from the Cr$^{3+}$-Te-Cr$^{3+}$ superexchange mechanism. Due to the weak intensity, broadness of the peak, and limitation of the spectral resolution of the detector, it would be erroneous to estimate the exact temperature at which the splitting occurs (see, **Fig. S6**), and splitting grows as magnetic order sets in near $T_C$. A strong SPC with a Cr$^{3+}$-Te-Cr$^{2+}$ configuration hinders the phonon breaking at low temperatures in TBA-CGT. Hence, it proves that the magnetic interaction changes from superexchange to double-exchange after intercalation. In addition to phonon splitting, we noticed dramatic changes in background Raman scattering at transition temperatures (see, **Fig. S6 and Fig. S7**). A large magnetic quasielectron scattering signal [10] again confirms an intercalated CGT's quasi-2D nature of magnetism.

**Conclusions**

By investigating the Raman fingerprints, we systematically studied the spin-phonon and electron-phonon coupling mechanisms and their close correlation with magnetic exchange interaction in an intercalated (quasi) 2D vdW magnetic TBA-CGT. Upon cooling, $E_g^3$ and $A_g^1$ Raman modes display noticeable changes in temperature consistent with the independent magnetization measurements. The magnetic and electronic coupling mechanisms are associated with the dramatic elevation of ferromagnetic Curie temperature and semiconductor-metal transition in the intercalated CGT. In addition, $E_g^4$ phonon mode has the largest SPC constant of about 3.03 cm$^{-1,}$ corresponding to the largest intraplanar Cr motions. Its temperature dependency successfully correlates with the existence of strong electron-phonon coupling during the electronic transition. We experimentally observe that the magnetic easy axis does not alter in CGT with $TBA^+$ intercalation, but the magnetic interaction changes from superexchange to double exchange through $Cr^{3+}$-Te-$Cr^{2+}$ links. This study presents a pathway to study tunable SPC dynamics in 2D magnetic material having numerous applications in spin filtering, spin Seebeck effect, spin wave control, etc.

**Acknowledgements**

This material is based upon work supported by the National Science Foundation Graduate Research Fellowship Program under Grant No. 184874.1 Any opinions, findings, and conclusions or recommendations expressed in this material are those of the author(s) and do not necessarily reflect the views of the National Science Foundation. S.R.S. and H.I. acknowledge support from the NSF-DMR (Award No. 2105109). SRS acknowledges support from NSF-MRI (Award No. 2018067). We acknowledge support of the Air Force Office of Scientific Research (AFOSR) Grant No. LRIR 23RXCOR003 and AOARD MOST Grant No. F4GGA21207H002 as well as general support from the Air Force Materials and Manufacturing (RX), Sensors (RY), and Aerospace Systems (RQ) Directorates. We also acknowledge support from the National Research Council's Senior NRC Associateship program sponsored by the National Academies of Sciences, Engineering, and Medicine.

**Figures:**

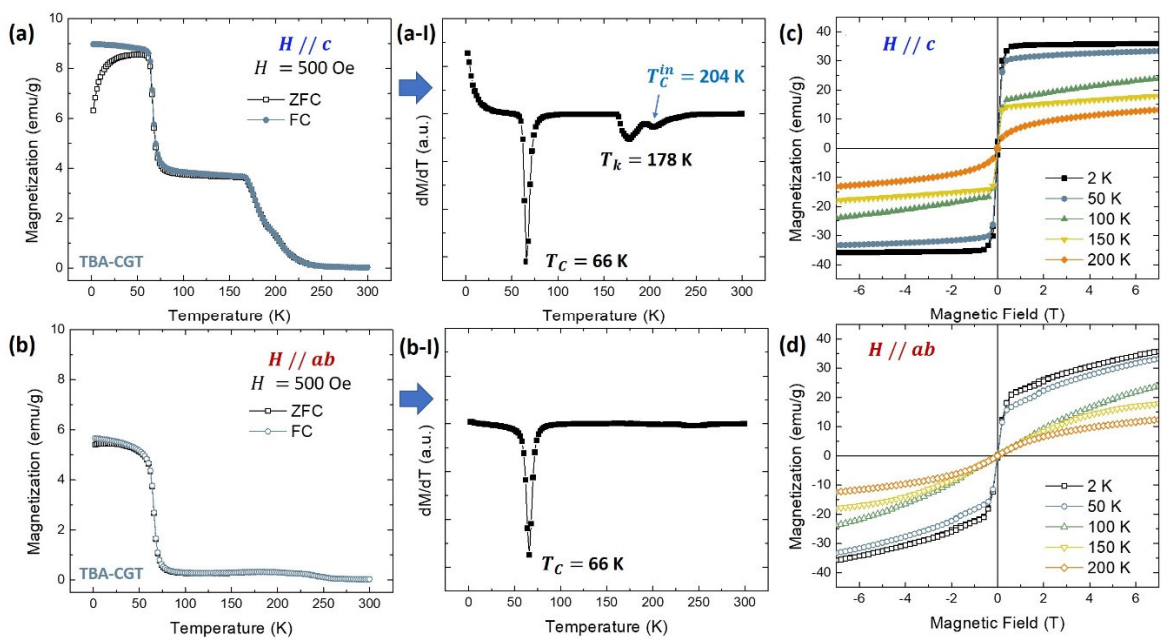

**Figure 1.** The temperature-dependent magnetization ($M - T$) collected on TBA-CGT with the magnetic field (a) $H//c$ and (b) $H//ab$ respectively. (a-I): The first derivative of M-T curve shown in figure a. (b-I): The first derivative of M-T curve shown in figure b. Magnetic hysteresis ($M - H$) loops collected at a few representative temperatures under (c) $H//c$ and (d) $H//ab$ configurations, respectively.

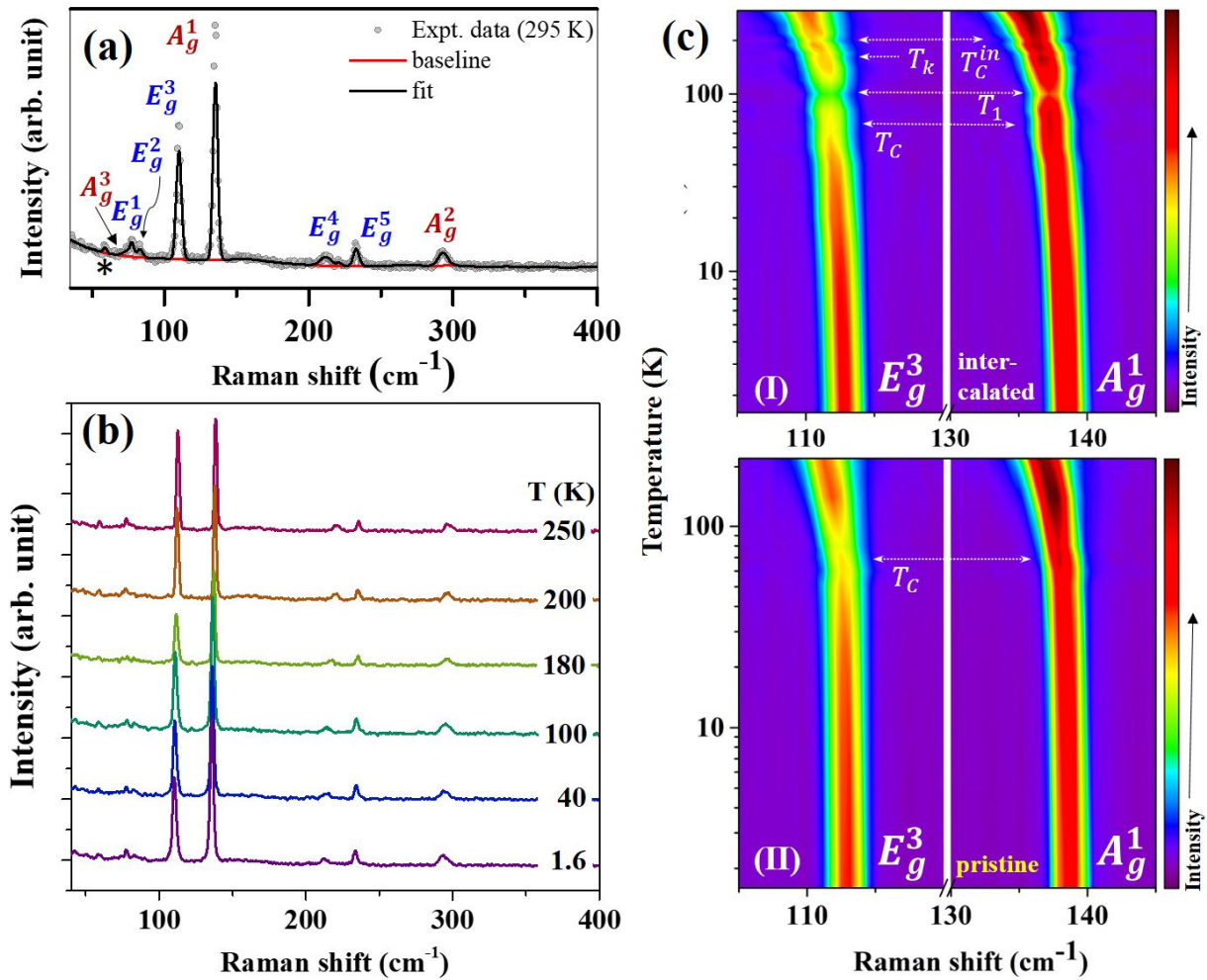

**Figure 2.** (a) Raman spectrum with the identified vibrational modes measured at the temperature of 295 K, and (b) the temperature evolution of Raman spectra measured from 1.6 – 250 K for TBA-CGT. A mode marked as "*" in (a) represents activated $A_{2u}$ mode (Raman inactive) and might be a part of the Davydov doublet. (c) 2D contour map of Raman spectra ($E_g^3$ and $A_g^1$ modes) for the (I) intercalated and (II) pristine CGT ranging from 1.6-295 K (see text).

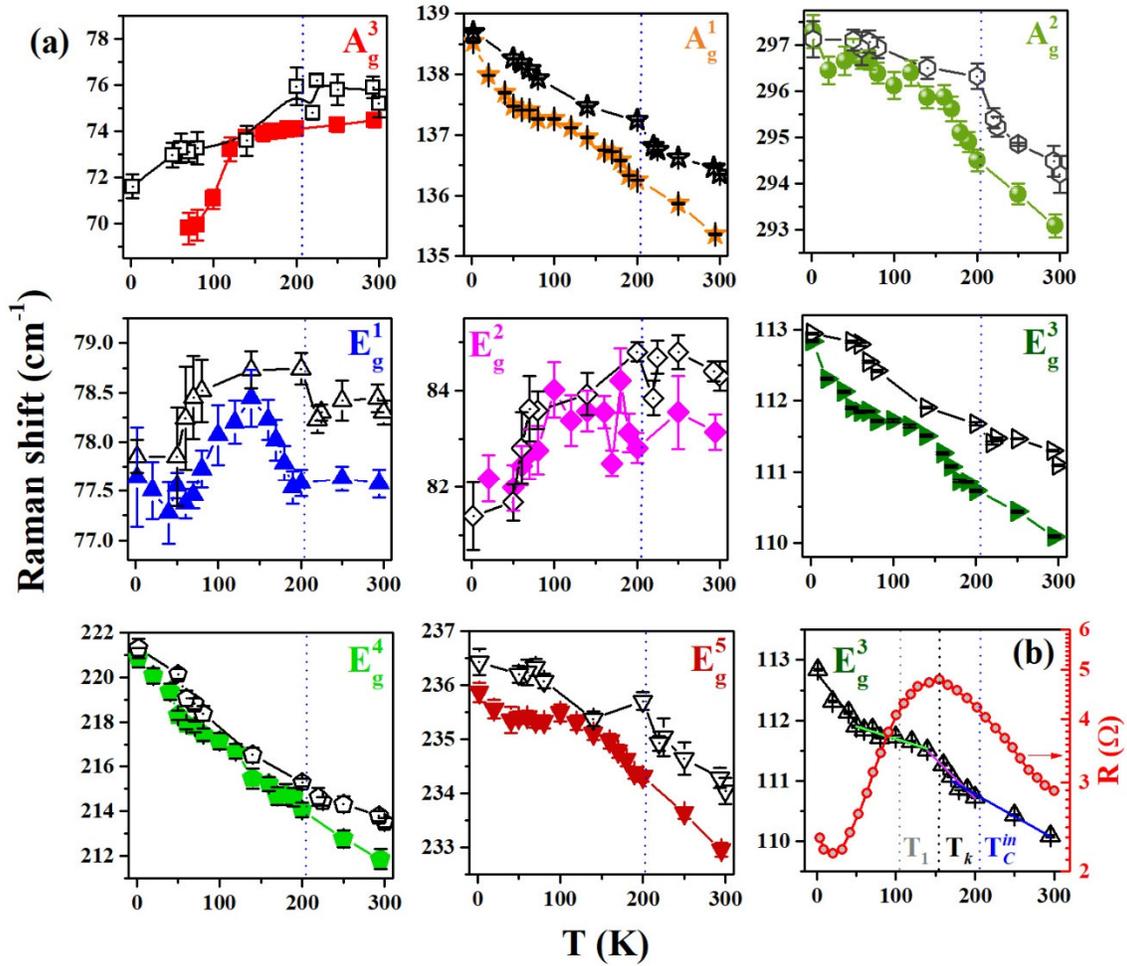

**Figure 3.** The temperature evolution of Raman vibrations (a) $A_g$ and $E_g$ modes in CGT (open symbols) and TBA-CGT (solid symbols). (b) Temperature dependency of the $E_g^3$ mode has been plotted with resistance for TBA-CGT to show the emergence of semiconductor-metal transition around $T_k$. The resistance values are extracted from the reference [6]. Solid lines depict the linear fits to the data (see text).

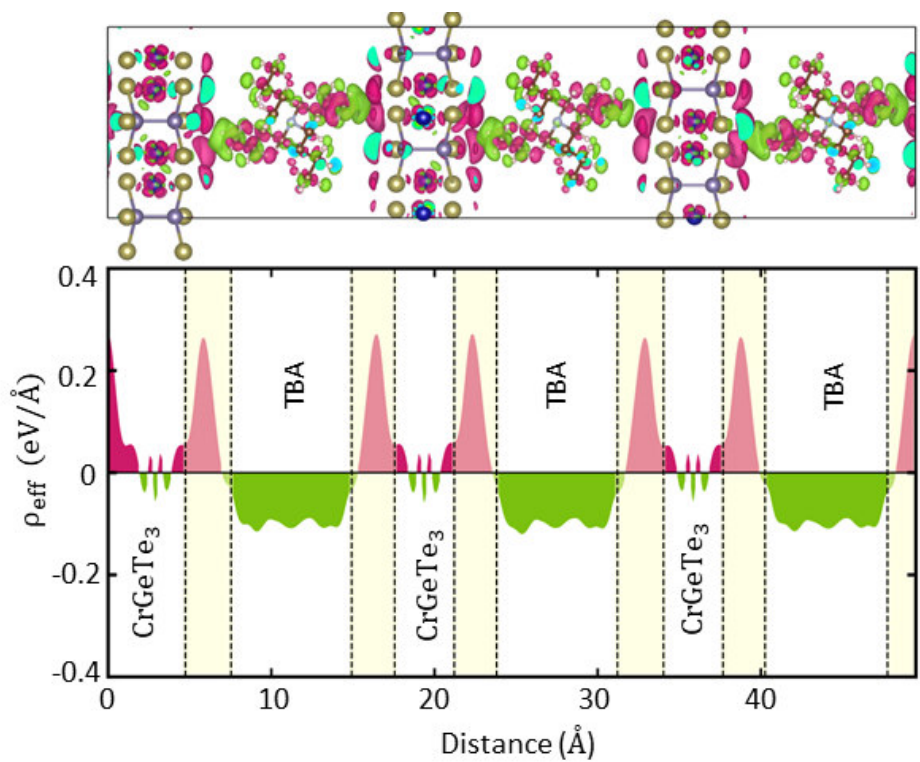

**Figure 4.** The iso-surface depicting the charge difference in real space is presented in the top panel, while the bottom panel displays the linear charge difference represented by $\rho_{eff}(z)$.

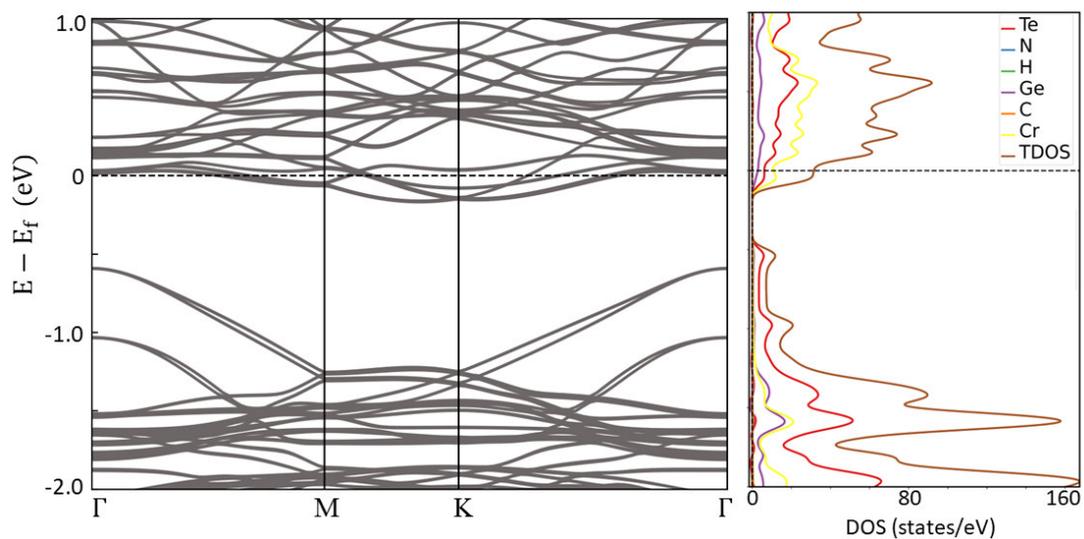

**Figure 5.** The electronic band structure over the complete Brillouin zone is shown on the left panel, and the corresponding orbital resolve density of the state is shown on the right panel.

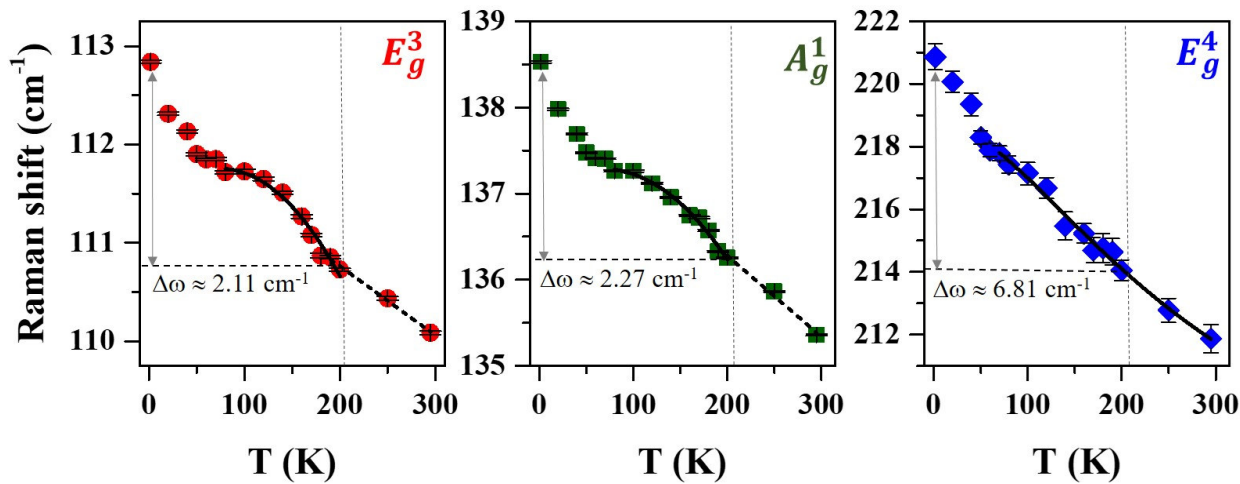

**Figure 6.** Temperature dependence of phonon frequency shifts for TBA-CGT. The solid line shows the model fits discussed in the texts. The vertical dash line indicates $T_c$.

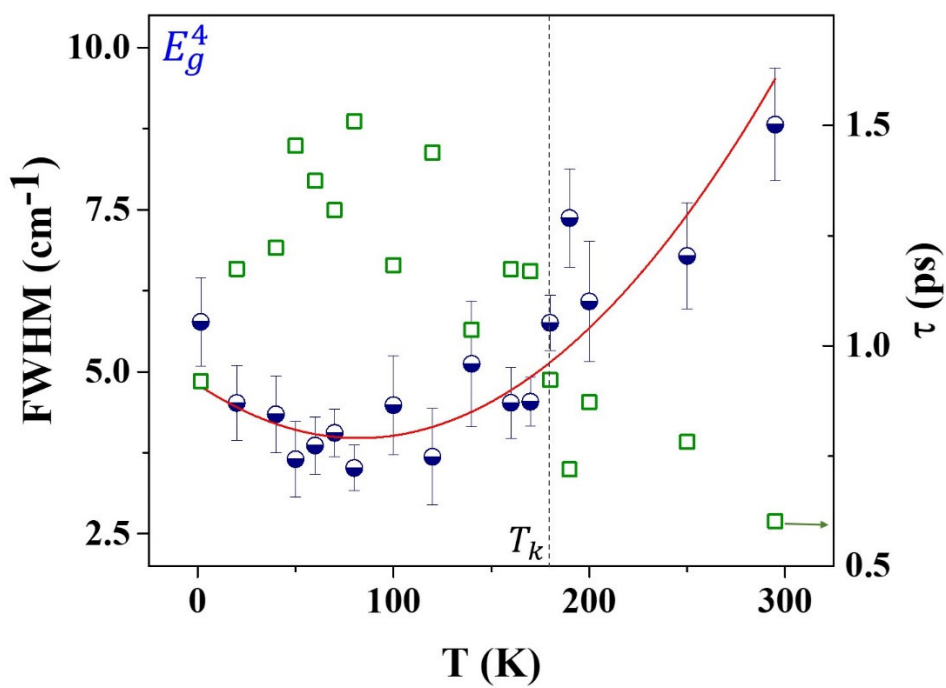

**Figure 7.** Temperature variation of FWHM (circles in blue) and phonon lifetime (open squares in green) for TBA-CGT have been plotted together for $E_g^4$ mode. The electron-phonon coupling, along with the three and four-phonon scattering terms, were fitted for $E_g^4$ mode (solid curve).

**Experimental methods:**

- **Sample synthesis**

We fabricated $Cr_2Ge_2Te_6$ using a flux growth reaction. We combined Cr powder (Alfa Aesar, ~20 mesh, 99.9%, reduced prior to use), Ge pieces (Alfa Aesar Puratronic, 99.9999+%), and Te pieces (Alfa Aesar Puratronic, 99.9999%) in the ratio of 2:6:36 in a 5 mL Canfield $Al_2O_3$ crucible set with an $Al_2O_3$ frit (LSP ceramics [18]). We sealed the crucibles in a quart ampoule with 1/3 atm Ar, placed the fixture in a large muffle furnace, ramped at 100°C/hr to 700°C, and held at this temperature for 300 hrs to ensure homogeneity. We then cooled to 500°C at a rate of 4°C/hr and decanted via centrifuge. The samples were micaceous and were maximum of 4 mm in diameter and ~ 40 μm in thickness. We recorded the discharge curve during the electrochemical discharge process during TBA+ intercalation (Fig. S2).

- **Sample characterization**

    **X-ray diffraction (XRD):** We took room temperature X-ray powder diffraction of the material's (00*l*) face with a Bruker D8 Discover system using Cu-$K_\alpha$ radiation (λ = 1.54056 Å) (Fig. S2). We performed compositional analysis via a Thermo Scientific Ultra Dry EDS spectrometer joined with a JEOL JSM-6060 SEM and found the composition to be $Cr_2Ge_2Te_6$, within instrument error.

    **X-ray photoelectron spectroscopy (XPS)**: XPS was performed using a Kratos Axis Ultra system with a monochromatic Al-$K_\alpha$ excitation source operating at 15 kV and 10 mA (Fig. S4). An intercalated CGT crystal was mounted on tape and pumped to a base pressure of < 2.0 x $10^{-9}$ Torr (2.7 x $10^{-7}$ Pa) before measurement. The detector was at a photoelectron take-off angle of 90° to the surface, and a pass energy of 20 eV was used. The native surface oxide layer of CGT

and TBA-CGT was removed using $Ar^+$ sputtering prior to photoelectron collecton. Data analysis was conducted with Shirley's background using CasaXPS.

**Magnetometry:** Low-temperature bulk magnetometry measurements for intercalated CGT were conducted using a Quantum Design Physical Property Measurement System with the AC Measurement System option. We extracted DC moment data in a temperature range extending from 5 K to 200 K and a maximum magnetic field of $\pm$ 30 kOe. Orientation-dependent measurements were conducted for both the in-plane (H//ab) and out-of-plane ($H//c$) direction of the magnetic field concerning the crystallographic $c$ −axis of the crystal.

**Raman spectroscopy:** Raman measurements were performed in the parallel polarization configuration on the bulk crystals. A fresh surface was exfoliated for the measurements. To ensure a reduction of noise in the signals, the samples were exfoliated. We conducted the measurements in the temperature range of 1.5 K to 295 K.

- **Theoretical calculations**

Electronic structure calculations were performed within the framework of density functional theory using the projector augmented wave method using the VASP package [19]. We use the generalized gradient approximation (GGA) method [20] with intra-site Hubbard $U$ for Cr d-electrons. We utilize a 450 eV cutoff energy, 3.9 eV Hubbard energy, and 0.58U Hund coupling on the $d$ −orbitals of the Cr-atoms [21, 22]. We have used the experimental parameters for our calculations, and the total energy was converged to $10^{-8}$ eV with a Gaussian-smearing method. We calculated using a 4×4×1 $\varGamma$ − centered $k$ − mesh with 16 $k$ − points in the Brillouin zone.

## SUPPORTING INFORMATION

**Raman fingerprints of spin-phonon coupling and magnetic transition in an organic molecule intercalated Cr₂Ge₂Te₆**

Sudeshna Samanta[1], Hector Iturriaga[1], Thuc T. Mai[2], Adam J. Biacchi[3], Rajbul Islam[4], Angela R. Hight Walker[2], Mohamed Fathi Sanad[5], Charudatta Phatak[6], Ryan Siebenaller[7,8], Emmanuel Rowe[7,9,10,11,12] Michael A. Susner[7], Fei Xue[4], Srinivasa R. Singamaneni[1]*

[1]Department of Physics, The University of Texas at El Paso, El Paso, Texas 79968, USA
[2]Quantum Metrology Division, Physical Measurement Laboratory, National Institute of Standards and Technology, Gaithersburg, Maryland 20899, USA
[3]Nanoscale Device Characterization Division, Physical Measurement Laboratory, National Institute of Standards and Technology, Gaithersburg, Maryland 20899, USA
[4]Department of Physics, University of Alabama at Birmingham, Birmingham, AL 35233
[5]Department of Chemical Engineering, Hampton University, Hampton, VA 23668, USA
[6]Materials Science Division, Argonne National Laboratory, Lemont, IL, USA.
[7]Materials and Manufacturing Directorate, Air Force Research Laboratory, Wright-Patterson Air Force Base, OH 45433 USA.
[8]Department of Materials Science and Engineering, The Ohio State University, Columbus, OH 43210 USA
[9]National Research Council, Washington, DC 20001, USA
[10]Department of Engineering Technology, Middle Tennessee State University, Murfreesboro, TN 37132, USA
[11]Department of Astronomy and Physics, Vanderbilt University, Nashville, TN 37235, USA
[12]Department of Life and Physical Sciences, Fisk University, Nashville, TN 37208, USA

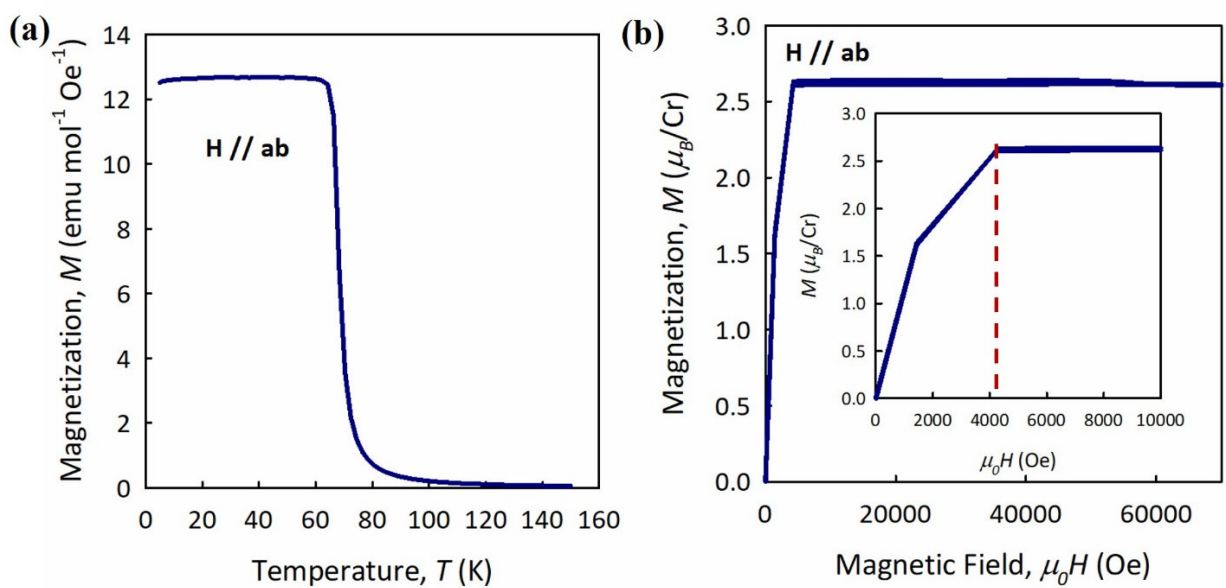

**Figure S1.** The temperature-dependent magnetization ($M - T$) and (b) field-dependent magnetization ($M - H$) with the magnetic field $H//ab$ collected on CGT. The Curie temperature for CGT is 66 K, and the saturation magnetic field is about 0.4 T for $H//ab$. The saturation magnetic moment is about 2.6 $\mu_B$/Cr atom for $H//ab$.

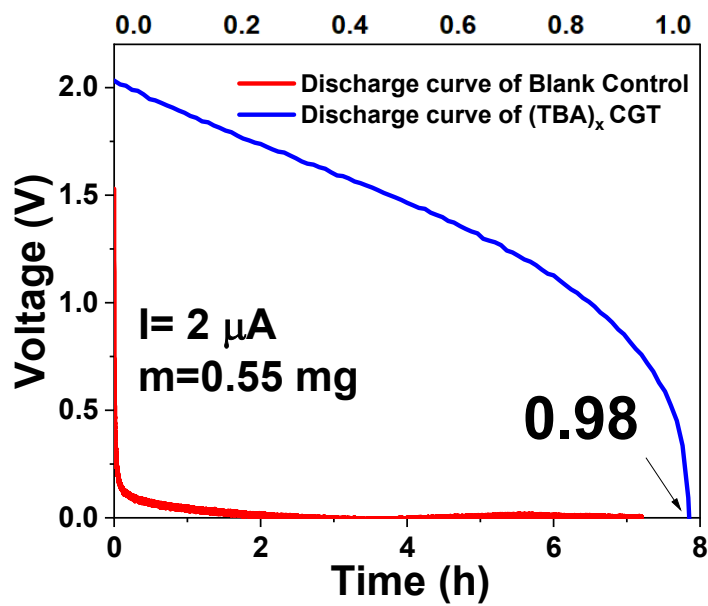

**Figure S2.** The discharge curve of (TBA)$_x$CGT during the electrochemical intercalation process.

The amount of intercalation ($x$) was calculated using experimental and theoretical methods. Figure S2 shows the characteristic discharge curve of CGT during the intercalation process. Control measurements were performed without using the CGT samples. With an applied current (2 μA) for the control sample, the voltage quickly drops where the electrolyte decomposes. In the electrochemical intercalation process, when $x > 0.98$, the voltage drops abruptly to 0V, indicating that the electrolyte starts to decompose. No further intercalation will happen beyond this point. Hence, from the discharge curve of (TBA)$_x$CGT, the maximum intercalation amount of TBA$^+$ in CGT would be ~0.98.

A pristine sample is a layered structure, and adjacent CGT layers are separated by an interlayer distance of 6.8 Å. In the presence of TBA intercalation, any in-plane slab shift is absent, and no subsequent change in in-plane lattice parameters was observed earlier [6]. The interlayer spacing increased to 16.48 Å with the inclusion of one layer of TBA$^+$ cations [6]. To reveal the potential channel of intercalation, we recorded the X-ray diffraction data for CGT before and after intercalation (Fig. S3(a)). We also confirm the increase of interlayer spacing in TBA-CGT by the shift of < 006 > peak in the lower angles in the X-ray diffraction data (Fig. S3(b)). It indicates that TBA$^+$ ions migrated along the spaces separated by < 006 > planes. It resembles the TBA$^+$ migration in $Fe_3GeTe_2$ crystals along the channels separated by < 002 > planes [23].

**X-ray diffraction:**

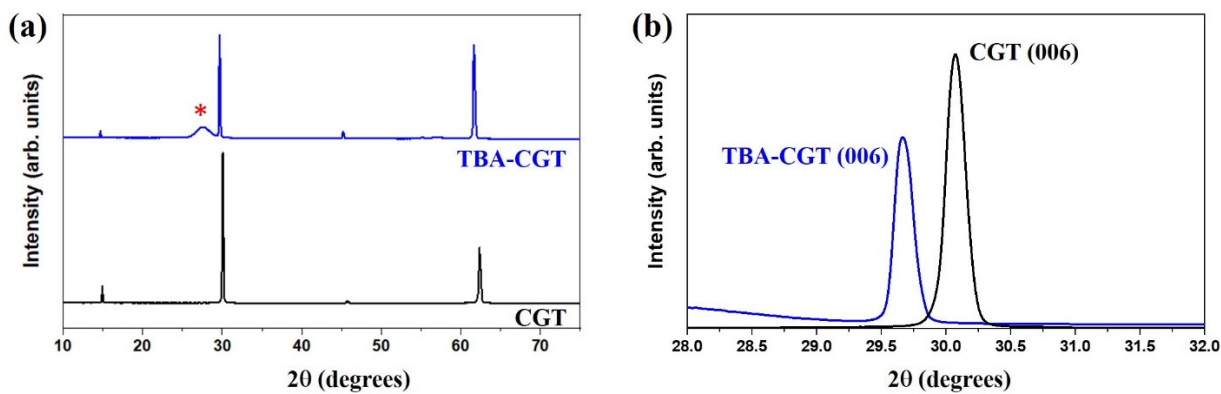

**Figure S3.** (a) X-ray diffraction pattern for CGT and intercalated TBA-CGT. The large hump around 27° $2\theta$ (asterisk) may be due to the TBA$^+$ causing a more considerable increase in the lattice spacing and probably significant strain in the system. (b) The intercalated TBA-CGT shows a shift of < 006 > peak to lower $2\theta$ angles w.r.t CGT. It confirms the increase in interplane distance relative to the pristine CGT and suggests successful intercalation.

The presence of $Cr^{3+}$, $Ge^{2+}$, and $Te^{2-}$ in CGT and TBA-CGT has been confirmed by sensitive X-ray photoelectron spectroscopy (XPS) (Fig. S4(a)-(f)). There is no trace of nitrogen observed in CGT (Fig. S4(c)), whereas, due to the high concentration of nitrogen in TBA-CGT, the corresponding XPS peak at 398.5 eV (Fig. S4(d)) confirms the successful intercalation path.

The amount of intercalated $TBA^+$ is calculated theoretically according to the following formula:

$$x = \frac{t * M * I}{F * m} \tag{1}$$

$t$ is the time of discharge during the intercalation process (7.7 h), $M$ is the molar mass of CGT, $I$ is the electric current passing through the cell, $F$ is the Faraday constant (96485.31 C mol$^{-1}$) and $m$ is the mass of pristine CGT single crystal. From the calculations, the amount of the $x$ was found to be 1.05, which is very close to the value calculated from the experimental method.

**X-ray photoelectron spectroscopy (XPS):**

Argon sputtering removed the surface oxide layers of TeO2 on crystals, and mixed valence states of $Ge^{4+}$ and $Ge^{2+}$ were observed. However, the Cr $2p$ − core level spectrum with $2p_{1/2}$ and $2p_{3/2}$ branches located at 587 and 578 eV, respectively, suggest the high-spin configuration of Cr ions ($Cr^{3+}$) in TBA-CGT. In addition, nitrogen is a light element and often becomes hard to detect in materials. There is no trace of nitrogen present in CGT (Fig. S4(c)), whereas, due to the high concentration of nitrogen in TBA-CGT, the corresponding XPS peak at 398.5 eV is highly visible (Fig. S4(d)). A negligible amount of solvent might be present and is responsible for the peak at ∼ 401 eV [23].

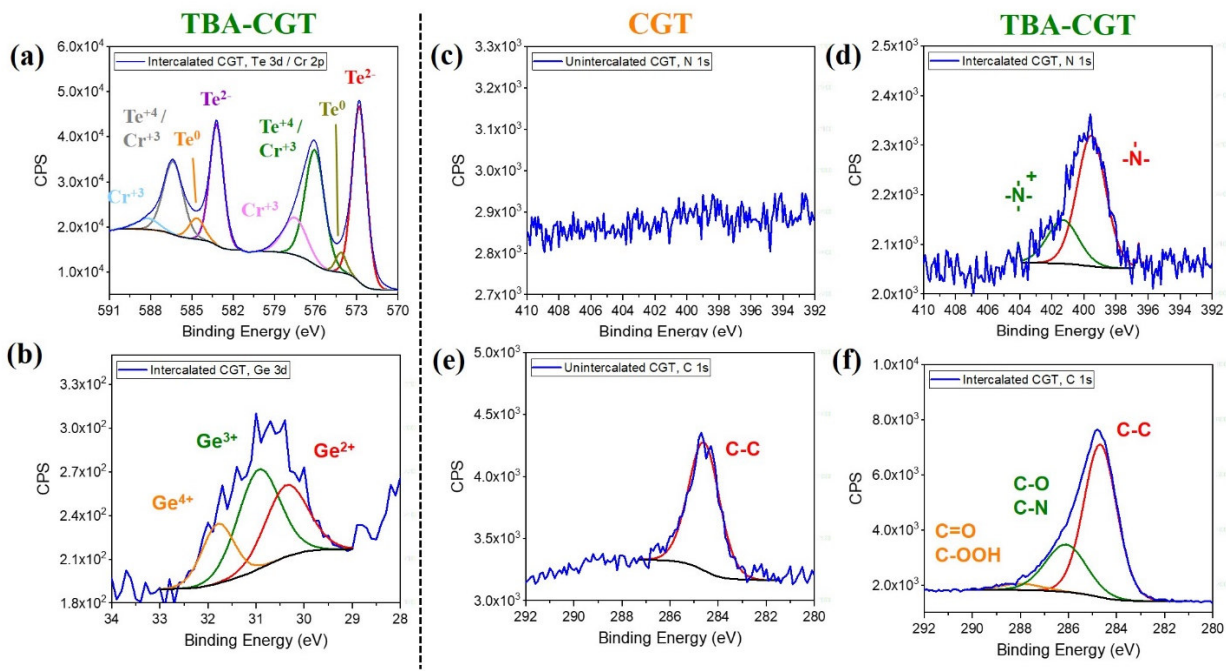

**Figure S4.** (a) - (b) Comparison of elementally resolved XPS data and deconvolution fitted curves of an intercalated $Cr_2Ge_2Te_6$ crystal with $TBA^+$ demonstrating the valencies of elements ($Cr^{3+}$, $Ge^{2+}$, and $Te^{2-}$). The nitrogen is (c) absent and (d) present in pristine and intercalated samples, respectively. It demonstrates the successful intercalation of $TBA^+$ in $Cr_2Ge_2Te_6$ crystals. (e) Only C-C bonds are seen in the unintercalated variant. (f) The intercalated CGT contains far more carbonaceous species and carbon bound to heteroatoms, which is absent in (e).

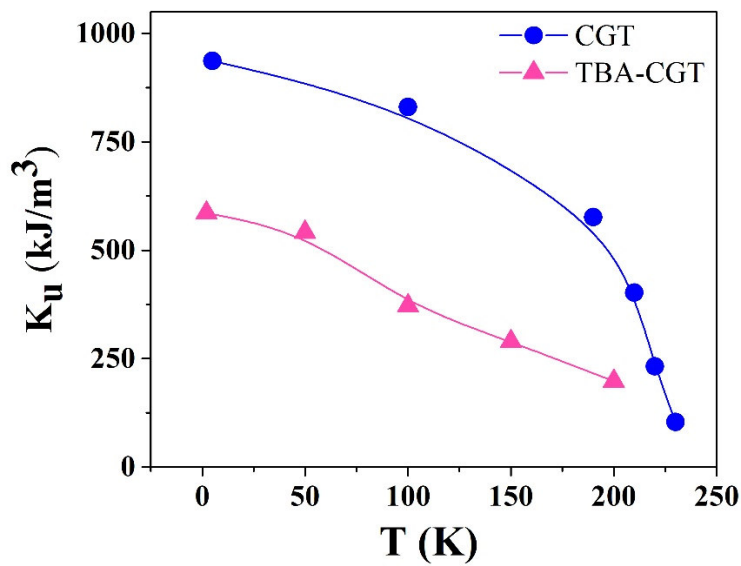

**Figure S5.** The uniaxial anisotropy coefficient is shown for both pristine (circle) and intercalated (triangle) in the low temperature region. The anisotropy parameter of CGT reduces significantly after intercalation.

**Table S-I: Raman modes for CGT and TBA-CGT at room temperature.**

| Sl. No. | Modes | CGT Raman shift (cm⁻¹) (Theory) | Assigned modes | CGT Raman shift (cm⁻¹) (Expt. from reference) | CGT Raman shift (cm⁻¹) (Expt.) | TBA-CGT Raman shift (cm⁻¹) (Expt.) |
|---|---|---|---|---|---|---|
| 1. | $A_{2u}$ (IR active) | 59.38 | | | 59.28 | 58.71 |
| 2. | $A_{2g}$ | 70.24 | $A_g^3$ | 74.8 | 75.20 | 74.48 |
| 3. | $E_g$ | 73.11 | $E_g^1$ | 78.6 | 78.30 | 77.59 |
| 4. | $E_g$ | 80.84 | $E_g^2$ | 85.3 | 84.33 | 83.15 |
| 5. | $A_{1g}$ | 104.75 | | | (indistinguishable) | (indistinguishable) |
| 6. | $E_g$ | 106.15 | $E_g^3$ | 110.1 | 111.13 | 110.08 |
| 7. | $A_{1g}$ | 130.06 | $A_g^1$ | 135.9 | 136.35 | 135.36 |
| 8. | $A_{2g}$ | 211.06 | | | (indistinguishable) | (indistinguishable) |
| 9. | $E_g$ | 211.08 | $E_g^4$ | 211.9 | 212.87 | 211.87 |
| 10. | $E_g$ | 226.01 | $E_g^5$ | 233.4 | 234.05 | 232.96 |
| 11. | $A_{1g}$ | 282.03 | $A_g^2$ | 292.6 | 294.2 | 293.08 |

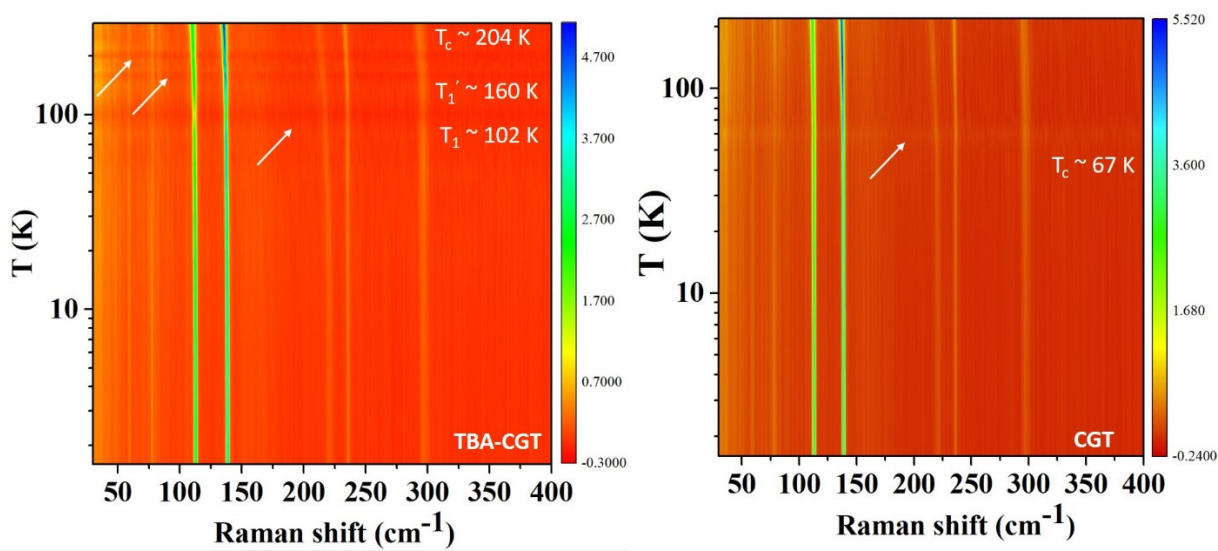

**Figure S6**. Temperature-dependent softening of $E_g^4$ Raman mode in TBA-CGT occurs with considerably higher rate (-0.034 cm$^{-1}$/K) than CGT (-0.011 cm$^{-1}$/K). The arrows indicate the spectral

changes in background signal due to quasi-electronic scattering near the magnetic and electric transitions.

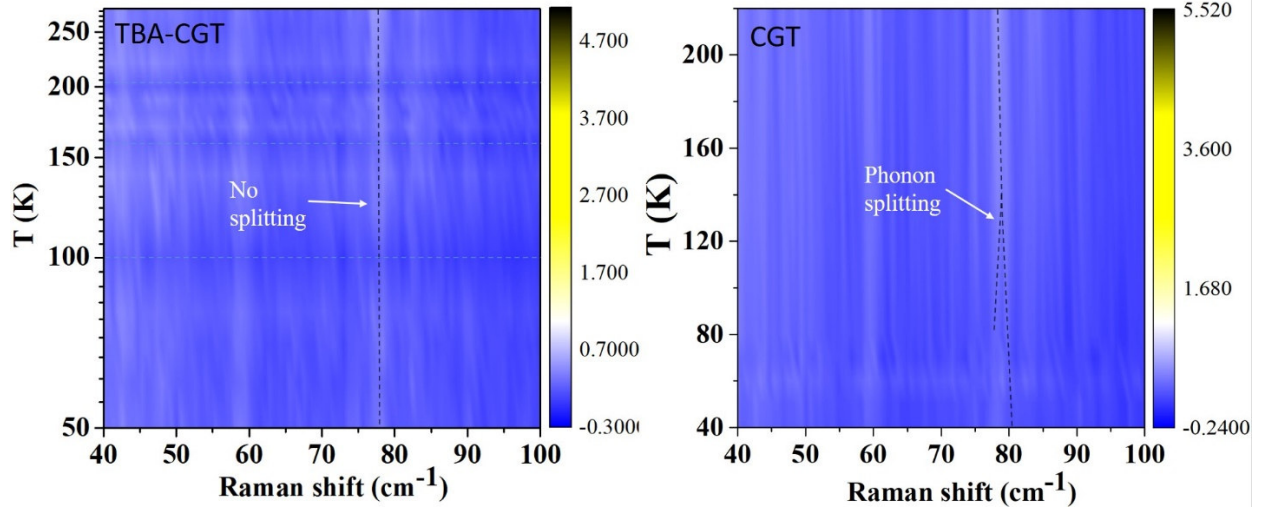

**Figure S7.** Phonon mode splitting is evident for the low-frequency $E_g^1$ mode in CGT, whereas the splitting is absent in TBA-CGT. The changes in Raman spectra background (horizontal lines in TBA-CGT) are prominent near electronic and magnetic transitions and correspond to the quasi-electronic scattering.

**Table S-II:**

| Modes | Frequency @ 200 K $\omega$ (cm⁻¹) | Frequency @ 0 K $\omega_0$ (cm⁻¹) | M₁ | M₂ | $\Delta \omega = \omega - \omega_0$ (cm⁻¹) | $\lambda' = \dfrac{\Delta \omega}{< S_i . S_j >}$ (cm⁻¹) (TBA-CGT) | $\lambda'$ (cm⁻¹) (CGT) [10] | $\lambda'$ (cm⁻¹) (CGT) (this study) |
|---|---|---|---|---|---|---|---|---|
| $E_g^3$ | 110.72 | 111.23 | 0.63 | -0.16 | 2.11 | 0.93 | 0.24 | 0.56 |
| | | ± 0.37 | ± 0.25 | ± 0.04 | | | | |
| $A_g^1$ | 136.25 | 136.85 ± 0.28 | 0.69 ± 0.23 | -0.22 ± 0.04 | 2.27 | 1.01 | 0.32 | 0.64 |
| $E_g^4$ | 214.05 | 222.10 ± 0.63 | - 4.09 ± 0.66 | 0.38 ± 0.14 | 6.81 | 3.03 | 1.2 | 2.77 |